\documentclass[12pt,a4paper]{article}
\usepackage{graphicx}
\usepackage[T1]{fontenc}
\usepackage[utf8]{inputenc}
\usepackage{textcomp}
\usepackage[sc,osf]{mathpazo}
\usepackage{a4wide}  
\usepackage{latexsym,amsthm,amsfonts,amsmath,mathrsfs,amssymb}
\usepackage{dsfont}
\usepackage{accents}
\usepackage[nosort]{cite}
\usepackage{booktabs} % for much better looking tables
\usepackage[unicode,implicit]{hyperref}
\hypersetup{%
  pdftitle    = {The first law of black hole mechanics in the Einstein-Maxwell
  theory revisited}
  pdfkeywords = {gauge, diffeomorphisms, Noether, Wald, action,
    Einstein-Maxwell, Reissner-Nordstrom, black hole, entropy, energy, charge,
  momentum map, thermodynamics, first law, zeroth law}
  pdfauthor   = {Zachary Elgood, Patrick Meessen and Tomas Ortin},
  plainpages  = true,
  colorlinks  = true,
  citecolor   = blue,
  urlcolor    = red,
  linkcolor   = black
}
\newcommand{\hepth}[1]{{\tt
\href{http://www.arXiv.org/abs/hep-th/#1}{hep-th/#1}}}
\newcommand{\grqc}[1]{{\tt
\href{http://www.arXiv.org/abs/gr-qc/#1}{gr-qc/#1}}}

\newcommand{\arxiv}[1]{{\tt arXiv:\href{http://www.arXiv.org/abs/#1}{#1}}}
\newcommand{\doi}[1]{{\tt DOI:\href{http://dx.doi.org/#1}{#1}}}

\makeatletter
\@addtoreset{equation}{section}
\makeatother

\begin{document}

\begin{flushright}
  \small
IFT-UAM/CSIC-20-068\\
\texttt{arXiv:2006.02792 [hep-th]}\\
May 28\textsuperscript{th}, 2020\\
\normalsize
\end{flushright}

\vspace{1cm}

\begin{center}

  {\Large {\bf The first law of black hole mechanics\\[.5cm] in the Einstein-Maxwell
  theory revisited}}

\vspace{1.5cm}

\renewcommand{\thefootnote}{\alph{footnote}}

{\sl\large Zachary Elgood,$^{1,}$}\footnote{Email: {\tt zachary.elgood[at]uam.es}}
{\sl\large Patrick Meessen$^{2,3,}$}\footnote{Email: {\tt meessenpatrick[at]uniovi.es}}
{\sl\large and Tom\'{a}s Ort\'{\i}n$^{1,}$}\footnote{Email: {\tt tomas.ortin[at]csic.es}}

\setcounter{footnote}{0}
\renewcommand{\thefootnote}{\arabic{footnote}}

\vspace{1.5cm}

${}^{1}${\it Instituto de F\'{\i}sica Te\'orica UAM/CSIC\\
C/ Nicol\'as Cabrera, 13--15,  C.U.~Cantoblanco, E-28049 Madrid, Spain}

\vspace{0.2cm}

${}^{2}${\it HEP Theory Group, Departamento de F\'{\i}sica, Universidad de Oviedo\\
  Avda.~Calvo Sotelo s/n, E-33007 Oviedo, Spain}\\

\vspace{0.2cm}

${}^{3}${\it Instituto Universitario de Ciencias y Tecnolog\'{\i}as Espaciales
  de Asturias (ICTEA)\\ Calle de la Independencia, 13, E-33004 Oviedo, Spain}\\

\vspace{1.5cm}

%%%%%%%%%%%%%%%%%%%%%%%%%%%%%%%%%%%%%%%%%%%%%%%%%%%%%%%%%%%%%%%%%%%%%%

{\bf Abstract}
\end{center}
\begin{quotation}
  {\small We re-derive the first law of black hole mechanics in the context of
    the Einstein-Maxwell theory in a gauge-invariant way introducing
    ``momentum maps'' associated to field strengths and the vectors that
    generate their symmetries. These objects play the role of generalized
    thermodynamical potentials in the first law and satisfy generalized zeroth
    laws, as first observed in the context of principal gauge bundles by
    Prabhu, but they can be generalized to more complex situations. We test
    our ideas on the $d$-dimensional Reissner-Nordstr\"om-Tangherlini black
    hole.}
\end{quotation}

\newpage
%%%%%%%%%%%%%%%%%%%%%%%%%%%%%%%%%%%%%%%%%%%%%%%%%%%%%%%%%%%%%%%%%%%%%%
%%%%%%%%%%%%%%%%%%%%%%%%%%%%%%%%%%%%%%%%%%%%%%%%%%%%%%%%%%%%%%%%%%%%%%
%%%%%%%%%%%%%%%%%%%%%%%%%%%%%%%%%%%%%%%%%%%%%%%%%%%%%%%%%%%%%%%%%%%%%%
%%%%%%%%%%%%%%%%%%%%%%%%%%%%%%%%%%%%%%%%%%%%%%%%%%%%%%%%%%%%%%%%%%%%%%
\pagestyle{plain}
%%%%%%%%%%%%%%%%%%%%%%%%%%%%%%%%%%%%%%%%%%%%%%%%%%%%%%%%%%%%%%%%%%%%%%
%%%%%%%%%%%%%%%%%%%%%%%%%%%%%%%%%%%%%%%%%%%%%%%%%%%%%%%%%%%%%%%%%%%%%%
%%%%%%%%%%%%%%%%%%%%%%%%%%%%%%%%%%%%%%%%%%%%%%%%%%%%%%%%%%%%%%%%%%%%%%
%%%%%%%%%%%%%%%%%%%%%%%%%%%%%%%%%%%%%%%%%%%%%%%%%%%%%%%%%%%%%%%%%%%%%%

\tableofcontents

%\newpage

%%%%%%%%%%%%%%%%%%%%%%%%%%%%%%%%%%%%%%%%%%%%%%%%%%%%%%%%%%%%%%%%%%%%%%
%%%%%%%%%%%%%%%%%%%%%%%%%%%%%%%%%%%%%%%%%%%%%%%%%%%%%%%%%%%%%%%%%%%%%%
%%%%%%%%%%%%%%%%%%%%%%%%%%%%%%%%%%%%%%%%%%%%%%%%%%%%%%%%%%%%%%%%%%%%%%
%%%%%%%%%%%%%%%%%%%%%%%%%%%%%%%%%%%%%%%%%%%%%%%%%%%%%%%%%%%%%%%%%%%%%%
\section*{Introduction}
%%%%%%%%%%%%%%%%%%%%%%%%%%%%%%%%%%%%%%%%%%%%%%%%%%%%%%%%%%%%%%%%%%%%%%
%%%%%%%%%%%%%%%%%%%%%%%%%%%%%%%%%%%%%%%%%%%%%%%%%%%%%%%%%%%%%%%%%%%%%%
%%%%%%%%%%%%%%%%%%%%%%%%%%%%%%%%%%%%%%%%%%%%%%%%%%%%%%%%%%%%%%%%%%%%%%
%%%%%%%%%%%%%%%%%%%%%%%%%%%%%%%%%%%%%%%%%%%%%%%%%%%%%%%%%%%%%%%%%%%%%%

Black-hole thermodynamics\footnote{For a recent review on black-hole
  thermodynamics with many references see, \textit{e.g.}~\cite{Wall:2018ydq}.}
is probably one of the most active fields of research in Theoretical
Physics. It interconnects seemingly disparate areas of Physics such as
Gravity, Quantum Field Theory and Information Theory providing deep insights
in all of them.

Black-hole thermodynamics originates in the analogy between the behaviour of
the area of the event horizon $A$ and the second law obeyed by the
thermodynamic entropy $S$ noticed by Bekenstein
\cite{Bekenstein:1972tm,Bekenstein:1973ur} in the results obtained by
Christodoulou and Hawking~\cite{Christodoulou:1970wf,Christodoulou:1972kt,Hawking:1971tu,Hawking:1971vc}. Shortly
afterwards, in Ref.~\cite{Bardeen:1973gs} Bardeen, Carter and Hawking extended
this analogy by proving another three laws of black hole mechanics similar to
the other three laws of thermodynamics involving the event horizon's surface
gravity $\kappa$ and angular velocity $\Omega$ and the black hole's mass $M$.
However, the analogy was only taken seriously after Hawking's discovery that
black holes radiate as black bodies with a temperature $T=\kappa/2\pi$
\cite{Hawking:1974sw}, which implied the relation $S=A/4$, both in
$c=G_{N}=\hbar=k=1$ units.

Ever since the formulation of these four laws, it has been tried to extend
their domain of application and validity with the inclusion of matter fields
and terms of higher-order in the curvature, for instance. In
Refs.~\cite{Lee:1990nz,Wald:1993nt,Iyer:1994ys} Wald and collaborators
developed a new approach to demonstrate the first law of black hole mechanics
in general diffeomorphism-invariant theories, beyond General Relativity. Since
the surface gravity relation to the Hawking temperature only depends on
generic properties of the event horizon, the quantity whose variation it
multiplies in the first law is naturally associated to the Bekenstein-Hawking
entropy $S$. This quantity, often called \textit{Wald entropy}, is just $A/4$
in General Relativity but, in more general theories, there can be additional
terms which can be understood, for instance, as $\alpha'$ corrections in
Superstring Theories
\cite{Cano:2018qev,Cano:2018brq,Edelstein:2019wzg,Cano:2019ycn,Elgood:2020xwu,Ortin:2020xdm}.

In the presence of matter fields, Wald's proof of the first law of black-hole
mechanics had to be re-examined because one of the main assumptions
Refs.~\cite{Wald:1993nt,Iyer:1994ys} is that all matter fields behave as
tensors and, simply put, there are no tensor fields in Nature apart form the
metric and scalar fields (if any); all of them have some sort of gauge
freedom and their transformations under diffeomorphisms are always coupled to
gauge transformations.  Indeed, as is well-known, fermionic fields coupled
to gravity transform under a local Lorentz group as spinors and bosonic fields
must transform under some gauge group if unwanted, typically negative-energy,
states are to be eliminated. The only scalar in the Standard Model, the Higgs
field, is, in fact, SU$(2)$ doublet.

The simplest matter field that, coupled to gravity, allows for black-hole
solutions is the Maxwell field \cite{kn:Reiss,kn:No8}. The presence of the
field introduces an additional term of the form $\Phi dQ$ in the first law
which takes into account the changes in the mass of the black hole when its
charge $Q$ changes. In this term $\Phi$ is the electric potential on the
horizon and a \textit{generalized zeroth law} states that it takes a constant
value over the horizon. The value of $\Phi$ is customarily taken to be
$k^{\mu}A_{\mu}$, where $k^{\mu}$ is the Killing vector for which the event
horizon is its associated Killing horizon and where it is assumed that the
electromagnetic field is in a gauge in which $\Phi$ is, indeed, constant.

This definition of $\Phi$ is clearly not gauge-invariant. This is a problem of
principle,\footnote{There are other problems as well: in Wald's approach, the
  Noether charge, which contains a term in which $\Phi$ occurs, is evaluated
  over the bifurcation surface, but the Maxwell field of the
  Reissner-Nordstr\"om black hole turns out to be singular there in the
  traditional gauge \cite{Gao:2003ys}.} which, as we are going to show, is
related to the more fundamental problem we were discussing: the fact that
Wald's proof of the first law does not deal properly with fields which have
some kind gauge freedom. In Wald's proof, one considers diffeomorphisms which
are symmetries of all the dynamical fields, but the naive definition of
invariance of fields with gauge freedom under diffeomorphsisms through the
standard Lie derivative is not gauge invariant.  This problem affects the
gravitational field itself when it is described in terms of the Vielbein
instead of the metric.

A solution for this particular case was provided in
Ref.~\cite{Jacobson:2015uqa} by defining the variation of the Vielbein under
diffeomorphisms through the Lie-Lorentz derivative
Refs.~\cite{kn:Lich,kn:Kos,kn:Kos2,Hurley:cf,Ortin:2002qb} which can be
understood as a generalization of the Lie derivative which transforms
covariantly under local Lorentz transformations. If the Vielbein is
annihilated by the Lie-Lorentz derivative with respect to some vector field in
some gauge it will be annihilated in any gauge and, as a matter of fact, the
vector field will be a Killing vector field of the metric. The Lie-Lorentz
derivative can be defined on all fields with Lorentz (spinor or vector)
indices, a fact that has been used to extend the proof of the first law of
black hole mechanics to supergravity in Ref.~\cite{Aneesh:2020fcr}.

A more general mathematically rigorous approach was proposed in
\cite{Prabhu:2015vua} using the formalism of principal gauge bundles which
encompasses Yang-Mills and Lorentz fields but, unfortunately, not the
Kalb-Ramond field or higher-rank form fields of string theory.\footnote{The
  first law has been proved for theories including one scalar and one $p$-form
  field in \cite{Compere:2007vx}, although the gauge-invariance problem has
  not been discussed in it.} Perhaps the most interesting result in that paper
is the realization that all the \textit{zeroth-laws} (the constancy of the
surface gravity, electric potential etc.) on the horizon fit into a common
pattern. In this paper we are going to recover and reformulate this result in
terms of the \textit{momentum map}, using gauge-covariant derivatives in which
this object plays a crucial role.\footnote{In
  Refs.~\cite{Frodden:2017qwh,Frodden:2019ylc}, which covers some of the
  topics studied here this object emerges as an ``improved gauge
  transformation''.}

Although gauge-covariant Lie derivatives are, perhaps, not the most
mathematically rigorous tool one can use, they can be generalized to
frameworks other than principal gauge bundles.\footnote{In this paper we will
  not consider those more complicated cases involving higher-rank $p$-form
  fields with Chern-Simons terms which typically arise in
  Superstring/Supergravity theories. We will consider the case of the
  Kalb-Ramond field with Yang-Mills and Lorentz Chern-Simons terms in its
  field strength in Ref.~\cite{kn:EMO}, where we will show how the
  gauge-covariant derivative approach with momentum maps that we introduce
  here provides a gauge-covariant, unambiguous results for the Wald-Noether
  charge.} Our goal in this paper is to show they can be consistently used in
a simpler context (the Einstein-Maxwell theory described in terms of
Vielbeins) and the objects to which the generalized zeroth law applies (here
the surface temperature and the electric potential) are the gauge-invariant
momentum maps associated to each gauge symmetry (Lorentz and U$(1)$) evaluated
over the horizon.

The emergence of the momentum map in this context may seem a bit strange; for
instance, there is no mention of it in Ref.~\cite{Jacobson:2015uqa} in spite
of their use of the (gauge-covariant) Lie-Lorentz derivative. However, as we
will show, the momentum map is indeed present in the Lie-Lorentz derivative
and plays the same role that the momentum map we will introduce for the
Maxwell case. As a matter of fact, gauge-covariant derivatives and the
momentum map arise most naturally in the study of superalgebras of symmetry,
when all the dynamical fields of a supergravity theory are left invariant by a
set of supersymmetry and bosonic transformations that combine diffeomorphisms,
gauge, local-Lorentz and local-supersymmetry transformations
\cite{Vandyck:ei,Vandyck:gc,Figueroa-O'Farrill:1999va,Ortin:2015hya}. This
object also plays a very interesting geometrical r\^ole in symmetric
Riemannian spaces and in certain spaces of special holonomy when they admit
Killing vectors that preserve their geometrical structures. When one wants to
gauge the corresponding symmetries in theories with $\sigma$-models of that
kind (typically supergravity theories) the momentum map plays an essential
role in the definition of the gauge-covariant derivative
\cite{Bandos:2016smv}.

This paper is organized as follows: in Section~\ref{sec-covariantLie} we
introduce the gauge-covariant derivatives that we are going to use:
Lie-Maxwell in Section~\ref{sec-LMderivatives} and Lie-Lorentz in
Section~\ref{sec-LieLorentz}. We also discuss the zeroth laws the respective
momentum maps obey. This last section is essentially a review of the
literature on the subject where we re-derive the formulae we are going to use
in the main body of the paper using our conventions (those of
Ref.~\cite{Ortin:2015hya}). In Section~\ref{sec-setup} we describe the
Einstein-Maxwell theory in $d$ dimensions (action and equations of motion) in
differential-form language and the $d$-dimensional
Reissner-Nordstr\"om-Tangherlini black hole solutions. In
Section~\ref{sec-WNcharge} we compute the Wald-Noether charge for this theory
using the transformations based on the gauge-covariant Lie derivatives defined
in Section~\ref{sec-covariantLie}. Then, in Section~\ref{sec-1stlaw} we prove
the first law for this system, identifying the Wald entropy, which we compute
for the Reissner-Nordstr\"om-Tangherlini black hole solutions. In
Section~\ref{sec-discussion} we briefly discuss our results and future
directions of research.

%%%%%%%%%%%%%%%%%%%%%%%%%%%%%%%%%%%%%%%%%%%%%%%%%%%%%%%%%%%%%%%%%%%%%%
%%%%%%%%%%%%%%%%%%%%%%%%%%%%%%%%%%%%%%%%%%%%%%%%%%%%%%%%%%%%%%%%%%%%%%
%%%%%%%%%%%%%%%%%%%%%%%%%%%%%%%%%%%%%%%%%%%%%%%%%%%%%%%%%%%%%%%%%%%%%%
%%%%%%%%%%%%%%%%%%%%%%%%%%%%%%%%%%%%%%%%%%%%%%%%%%%%%%%%%%%%%%%%%%%%%%
\section{Covariant Lie derivatives and momentum maps}
\label{sec-covariantLie}
%%%%%%%%%%%%%%%%%%%%%%%%%%%%%%%%%%%%%%%%%%%%%%%%%%%%%%%%%%%%%%%%%%%%%%
%%%%%%%%%%%%%%%%%%%%%%%%%%%%%%%%%%%%%%%%%%%%%%%%%%%%%%%%%%%%%%%%%%%%%%
%%%%%%%%%%%%%%%%%%%%%%%%%%%%%%%%%%%%%%%%%%%%%%%%%%%%%%%%%%%%%%%%%%%%%%
%%%%%%%%%%%%%%%%%%%%%%%%%%%%%%%%%%%%%%%%%%%%%%%%%%%%%%%%%%%%%%%%%%%%%%

One of the main ingredients in the proofs of the first law of black hole
mechanics using Wald's formalism \cite{Wald:1993nt,Iyer:1994ys} is the use of
infinitesimal diffeomorphisms that leave invariant all the dynamical fields.

If we use the metric $g_{\mu\nu}$ as dynamical field, since the metric is just
a tensor, its transformation under infinitesimal diffeomorphisms
$\delta_{\xi}x^{\mu} =\xi^{\mu}(x)$ is given by (minus) the standard Lie
derivative

\begin{equation}
  \delta_{\xi} g_{\mu\nu}
  =
  -\pounds_{\xi} g_{\mu\nu} = -2\nabla_{(\mu}\xi_{\nu)}\, ,
\end{equation}

\noindent
which vanishes when $\xi^{\mu}$ is a Killing vector of $g_{\mu\nu}$. We will
distinguish Killing vectors from generic vectors $\xi^{\mu}$ denoting them by
$k^{\mu}$.

If, as we want to do here, we use as dynamical field the Vielbein
$e^{a}{}_{\mu}$ instead of $g_{\mu\nu}$, in order to define its symmetries, we
face the well-known problem of the gauge freedom of $e^{a}{}_{\mu}$, which in
this context has been treated in Refs.~\cite{Jacobson:2015uqa,Prabhu:2015vua}.
The same happens with the electromagnetic potential $A_{\mu}$, which also has
been treated in this context in Refs.~\cite{Prabhu:2015vua}.

One way to deal with this problem is to define a gauge-covariant notion of Lie
derivative. The Lie derivative in the corresponding principal bundle, used in
Ref.~\cite{Prabhu:2015vua} provides the most rigorous definition such a
derivative. Here we will introduce a less sophisticated version that makes use
of the so-called \textit{momentum map} and which can be defined for more
general fields such as the Kalb-Ramond 2-form of the Heterotic Superstring,
which cannot be described in the framework of a principal bundle
\cite{kn:EMO}.  Gauge-covariant derivatives arise naturally in the commutator
of two local supersymmetry transformations and in the construction of Lie
superalgebras of supersymmetric backgrounds
\cite{Vandyck:ei,Vandyck:gc,Figueroa-O'Farrill:1999va,Ortin:2015hya}.

Due to its simplicity, we start with the Maxwell field.

%%%%%%%%%%%%%%%%%%%%%%%%%%%%%%%%%%%%%%%%%%%%%%%%%%%%%%%%%%%%%%%%%%%%%%
%%%%%%%%%%%%%%%%%%%%%%%%%%%%%%%%%%%%%%%%%%%%%%%%%%%%%%%%%%%%%%%%%%%%%%
%%%%%%%%%%%%%%%%%%%%%%%%%%%%%%%%%%%%%%%%%%%%%%%%%%%%%%%%%%%%%%%%%%%%%%
%%%%%%%%%%%%%%%%%%%%%%%%%%%%%%%%%%%%%%%%%%%%%%%%%%%%%%%%%%%%%%%%%%%%%%
\subsection{Lie-Maxwell derivatives}
\label{sec-LMderivatives}
%%%%%%%%%%%%%%%%%%%%%%%%%%%%%%%%%%%%%%%%%%%%%%%%%%%%%%%%%%%%%%%%%%%%%%
%%%%%%%%%%%%%%%%%%%%%%%%%%%%%%%%%%%%%%%%%%%%%%%%%%%%%%%%%%%%%%%%%%%%%%
%%%%%%%%%%%%%%%%%%%%%%%%%%%%%%%%%%%%%%%%%%%%%%%%%%%%%%%%%%%%%%%%%%%%%%
%%%%%%%%%%%%%%%%%%%%%%%%%%%%%%%%%%%%%%%%%%%%%%%%%%%%%%%%%%%%%%%%%%%%%%

The electromagnetic field $A_{\mu}$ is a field with gauge freedom: we must
consider physically equivalent two configurations that are related by the
gauge transformation

\begin{equation}
  \delta_{\chi}A_{\mu} = \partial_{\mu}\chi\, ,
\end{equation}

\noindent
and, furthermore, as a general rule, it is not possible to give a globally
regular expression of the electromagnetic field in a single
gauge.\footnote{The main example of this situation is the magnetic monopole
  \cite{Wu:1975es}.} However, the standard Lie derivative does not commute
with these gauge transformations and gives different results in different
gauges. This is why a gauge-covariant notion of Lie derivative is needed in
this case.

In the subsequent discussion it is convenient to use differential-form
language. In terms of the electromagnetic 1-form potential
$A\equiv A_{\mu}dx^{\mu}$, we define the electromagnetic field strength 2-form
by $F=dA$ so that it satisfies the Bianchi identity $dF=0$. In components we
have

\begin{equation}
  \label{eq:Fmn}
  F \equiv \tfrac{1}{2}F_{\mu\nu}dx^{\mu}\wedge dx^{\nu}\, ,
  \hspace{1cm}
  F_{\mu\nu} = 2\partial_{[\mu}A_{\nu]}\, .
\end{equation}

The field strength is invariant under the gauge transformations
$\delta_{\chi}A=d\chi$ and we can treat it as a standard 2-form whose
transformation under infinitesimal diffeomorphisms generated by $\xi^{\mu}$ is
given by (minus) the standard Lie derivative which, on $p$-forms, acts as
$\pounds_{\xi} =\imath_{\xi}d +d\imath_{\xi}$.\footnote{In our conventions,
  for a $p$-form $\omega^{(p)}$ with components
  $\omega^{(p)}{}_{\mu_{1}\cdots \mu_{p}}$, $\imath_{\xi}\omega^{(p)}$ is the
  $(p-1)$-form with components
  $(\imath_{\xi}\omega^{(p)})_{\mu_{1}\cdot\mu_{p-1}}=\xi^{\nu}\omega^{(p)}{}_{\nu\mu_{1}\cdot\mu_{p-1}}$.}

Using the Bianchi identity we find that

\begin{equation}
\delta_{\xi}F = -d\imath_{\xi}F\, .
\end{equation}

If $\xi$ is a symmetry of all the dynamical fields, in which case we will
denote it by $k$, we have that $\delta_{k}F=0$ and the above equation implies
that, locally, there is a gauge-invariant function $P_{k}$ called
\textit{momentum map} such that\footnote{The sign of $P_{k}$ is purely
  conventional.}

\begin{equation}
  \label{eq:Pkdef}
\imath_{k}F = -dP_{k}\, .
\end{equation}

\noindent
$P_{k}$ is defined by this equation up to an additive constant that we will
discuss later.

Let us now consider the variation of $A$ under infinitesimal diffeomorphisms,
which, according to general arguments (see
\textit{e.g.}~Refs.~\cite{Ortin:2015hya,Prabhu:2015vua}) has to be given
locally by a combination of (minus) the Lie derivative and a ``compensating''
gauge transformation generated by a $\xi$-dependent parameter $\chi_{\xi}$
which is to be determined by demanding that $\delta_{k}A=0$ when
$\delta_{k}F=0$:

\begin{equation}
  \delta_{\xi}A
  =
  -\pounds_{\xi}A +d\chi_{\xi}
  =
  -\imath_{\xi}F
+d\left(\chi_{\xi}-\imath_{\xi}A\right)\, .  
\end{equation}

Then, taking into account Eq.~(\ref{eq:Pkdef}), we conclude that

\begin{equation}
  \label{eq:LMparameter}
\chi_{\xi}= \imath_{\xi}A -P_{\xi}\, ,
\end{equation}

\noindent
where $P_{\xi}$ is a function of $\xi$ which satisfies Eq.~(\ref{eq:Pkdef})
when $\xi=k$ and generates a symmetry of all the dynamical fields.

It is natural to identify the above transformation $ \delta_{\xi}A$ with
(minus) a gauge-covariant Lie derivative of $A$ that we can call
\textit{Lie-Maxwell derivative}

\begin{equation}
  \label{eq:LieMaxwelldef}
  \delta_{\xi}A = - \mathbb{L}_{\xi}A\, ,
  \hspace{1.5cm}
  \mathbb{L}_{\xi}A \equiv \imath_{\xi}F +dP_{\xi}\, .
\end{equation}

While this derivative does not enjoy the most important property of Lie
derivatives $[\pounds_{\xi},\pounds_{\eta}] = \pounds_{[\xi,\eta]}$ for
generic vector fields $\xi,\eta$, it is clear that it does for those that
generate symmetries of $A$ and $F$ and annihilates them. This is certainly
enough for us.

For stationary asymptotically-flat black holes, when the Killing vector $k$ is
the one normal to the event horizon, the momentum map can be understood as the
electric potential $\Phi$ which, evaluated on the horizon
$\Phi_{\mathcal{H}}$, appears in the first law.\footnote{See, for instance
  Ref.~\cite{Kunduri:2013vka} for a proof of the first law in the context of
  5-dimensional supergravity and the role that $\Phi$ plays in it.} In the
early literature (see \textit{e.g.}~Section~6.3.5 of
Ref.~\cite{Frolov:1998wf}) it was assumed from the start that there is a gauge
in which

\begin{equation}
  \pounds_{k}A = \imath_{k}dA+d(\imath_{k}A)
  =
  0\, .
\end{equation}

\noindent
Then, the electric potential $\Phi$ was identified with $\imath_{k}A$ because,
according to the above equation, $d\Phi=-\imath_{k}F$, which can be defined as the
electric field for an observer associated to the time direction defined by
$k$.

It is clear that $P_{k}$ can be identified with $\Phi$ (both satisfy the same
equation). However, in a general gauge, it will not be given by just $\imath_{k}A$
and we will have to compute it. Nevertheless, the main property of $\Phi$,
namely the fact that it is constant over the horizon (sometimes called
\textit{generalized zeroth law}) still holds because it is, actually, a
property of $-\imath_{k}F$ based on the properties of $k$, the Einstein equations
and the assumption that the energy-momentum tensor of the electromagnetic
field satisfies the dominant energy condition.

%%%%%%%%%%%%%%%%%%%%%%%%%%%%%%%%%%%%%%%%%%%%%%%%%%%%%%%%%%%%%%%%%%%%%%
%%%%%%%%%%%%%%%%%%%%%%%%%%%%%%%%%%%%%%%%%%%%%%%%%%%%%%%%%%%%%%%%%%%%%%
%%%%%%%%%%%%%%%%%%%%%%%%%%%%%%%%%%%%%%%%%%%%%%%%%%%%%%%%%%%%%%%%%%%%%%
%%%%%%%%%%%%%%%%%%%%%%%%%%%%%%%%%%%%%%%%%%%%%%%%%%%%%%%%%%%%%%%%%%%%%%
\subsection{Lie-Lorentz derivatives}
\label{sec-LieLorentz}
%%%%%%%%%%%%%%%%%%%%%%%%%%%%%%%%%%%%%%%%%%%%%%%%%%%%%%%%%%%%%%%%%%%%%%
%%%%%%%%%%%%%%%%%%%%%%%%%%%%%%%%%%%%%%%%%%%%%%%%%%%%%%%%%%%%%%%%%%%%%%
%%%%%%%%%%%%%%%%%%%%%%%%%%%%%%%%%%%%%%%%%%%%%%%%%%%%%%%%%%%%%%%%%%%%%%
%%%%%%%%%%%%%%%%%%%%%%%%%%%%%%%%%%%%%%%%%%%%%%%%%%%%%%%%%%%%%%%%%%%%%%

The original motivation for the definition of a derivative covariant under
local Lorentz transformations, often called the Lie-Lorentz derivative, was
its need for the proper treatment of spinorial fields in curved spaces in such
a way that the flat-space results were correctly recovered.

In Minkowski spacetime, fermionic fields transform in spinorial
representations of the Lorentz group, which leaves invariant the spacetime
metric $(\eta_{ab})=\text{diag}(+-\cdots -)$. Since generic spacetime metrics
$g_{\mu\nu}$ do not have any isometries, the Lorentz group will not be
realized as a group of general coordinate transformations (g.c.t.s) leaving
invariant the spacetime metric. Weyl realized that, if one introduces an
orthonormal base in cotangent space at a given point in spacetime

\begin{equation}
\{e^{a}=e^{a}{}_{\mu}dx^{\mu}\}\, ,
\hspace{1cm}
e^{a}{}_{\mu}e^{b}{}_{\nu}g^{\mu\nu}=\eta^{ab}\, ,
\end{equation}

\noindent
the Lorentz group arises naturally as the group of linear transformations of
the base

\begin{equation}
  e^{a\, \prime}
  =
  \Lambda^{a}{}_{b}e^{b} \sim  (\eta^{a}{}_{b} +\sigma^{a}{}_{b})e^{b}\, ,
\end{equation}

\noindent
($\sigma^{a}{}_{b}$ are the infinitesimal transformations) that
preserves orthonormality.

\begin{equation}
  \Lambda^{a}{}_{c}\Lambda^{b}{}_{d}\eta^{cd}=\eta^{ab}\, ,
  \,\,\,\,\,
  \Rightarrow
  \,\,\,\,\,
  \sigma^{(a}{}_{c}\eta^{b)c} = \sigma^{(ab)} = 0\, .
\end{equation}

In Ref.~\cite{kn:Weyl4}, Weyl proposed to define fermionic fields
$\psi$ as fields transforming in the spinorial representation of the
Lorentz group that acts in the tangent and cotangent space, that is

\begin{equation}
\delta_{\sigma}\psi \equiv \tfrac{1}{2}\sigma^{ab}\Gamma_{s}(M_{ab})\psi\, , 
\end{equation}

\noindent
where $\Gamma_{r}(M_{ab})$ stands for the matrices that represent the
generators of the Lorentz group $\{M_{ab}\}$ in the representation
$r$. As is well-known, the generators in the spinorial
representation can be constructed taking antisymmetrized products of
the gamma matrices $\gamma^{a}$, $\gamma^{ab}\equiv \gamma^{[a}\gamma^{b]}$

\begin{equation}
\Gamma_{s}(M_{ab}) = \tfrac{1}{2}\gamma_{ab}\, ,  
  \,\,\,\,\,
  \Rightarrow
  \,\,\,\,\,
\delta_{\sigma}\psi \equiv \tfrac{1}{4}\sigma^{ab}\gamma_{ab}\psi\, . 
\end{equation}

Since these transformations can be different at each point, the
Lorentz parameters $\sigma^{ab}$ take different values at different
points of the spacetime and become functions $\sigma^{ab}(x)$ which
will be smooth if the bases of the tangent and cotangent space are
assumed to vary smoothly so that they are smooth vector and 1-form
fields.

Theories containing fermionic fields in curved spacetimes are required to be
invariant under these local Lorentz transformations. Their construction
demands the introduction of a gauge field, the so-called spin connection
1-form, conventionally denoted by $\omega^{ab}=\omega_{\mu}{}^{ab}
dx^{\mu}$. The spin connection enters the Lorentz-covariant derivatives of any
field $T$ (indices not shown) transforming in the representation $r$ of the
Lorentz group as follows:

\begin{equation}
  \mathcal{D} T^{(r)}
  \equiv
  \left[d -\tfrac{1}{2}\omega^{ab}\Gamma_{r}(M_{ab}) \right]T^{(r)}\, .  
\end{equation}

\noindent
The transformation properties of $T^{(r)}$ are preserved by the covariant
derivative if, under infinitesimal local Lorentz transformations,

\begin{equation}
  \label{eq:deltasigmaomega}
  \delta_{\sigma}\omega^{ab}
  =
  \mathcal{D}\sigma^{ab}
  =
  \left[d
    -\tfrac{1}{2}\omega^{cd}\Gamma_{Adj}(M_{cd})\right]\sigma^{ab}
    =
    d\sigma^{ab}-2\omega^{[a}{}_{c}\sigma^{|c|b]}\, .
\end{equation}

From now on $\nabla_{\mu}$ will denote the full (affine plus Lorentz)
covariant derivative satisfying the first Vielbein postulate

\begin{equation}
  0=\nabla_{\mu}e^{a}{}_{\nu}
  \equiv 
  \partial_{\mu}e^{a}{}_{\nu} -\omega_{\mu}{}^{a}{}_{b}e^{b}{}_{\nu}
  -\Gamma_{\mu\nu}{}^{\rho}e^{a}{}_{\rho}\, .
\end{equation}

\noindent
On pure Lorentz tensors $\nabla=\mathcal{D}$.

Now, how do spinors and general Lorentz tensors transform under
infinitesimal g.c.t.s generated by an vector field $\xi$?

Customarily, these fields are treated as scalars, so that, if
$\pounds_{\xi}$ stands for the standard Lie derivative,

\begin{equation}
  \label{eq:naivegct}
\delta_{\xi}T = -\pounds_{\xi}T = -\imath_{\xi}dT\, . 
\end{equation}

There are many reasons why this has to be wrong. For starters, if we
consider the particular case of a vector field $\xi$ generating a global
Lorentz transformation in Minkowski spacetime
$\xi^{\mu} = \sigma^{\mu}{}_{\nu}x^{\nu}+a^{\mu}$, the transformation in
Eq.~(\ref{eq:naivegct}) is completely different from the transformation of a
Lorentz tensor

\begin{equation}
\delta_{\sigma}T = \tfrac{1}{2}\sigma^{ab}\Gamma_{r}(M_{ab})T\, .
\end{equation}

\noindent
However, it should reduce to this if the Fermionic fields introduced in curved
spacetimes via Weyl's prescription have anything to do with the standard
special-relativistic Fermionic fields.

Furthermore, it is clear that the effect of the g.c.t.~Eq.~(\ref{eq:naivegct})
on $T$ depends on the gauge, or, equivalently, on the choice of tangent space
basis. In other words the expression for $\delta_{\xi}$ in
Eq.~(\ref{eq:naivegct}) is not covariant under local Lorentz transformations.

Indeed, Lorentz tensors are not scalar nor tensor fields under g.c.t.s.  They
are sections of some bundle or, at a more pedestrian level, they are fields
that, under g.c.t.s, transform as world tensors up to a local Lorentz
transformation whose parameter depends on the field and on the generator of
the g.c.t. $\sigma^{ab}_{\xi}$.

Then, instead of Eq.~(\ref{eq:naivegct}) we must write

\begin{equation}
  \delta_{\xi}T
  =
  -\pounds_{\xi}T+\delta_{\sigma_{\xi}}T\, ,
\end{equation}

\noindent
where $\sigma_{\xi}{}^{ab}$ makes $\delta_{\xi}T$ covariant under further
local Lorentz transformations.

The parameter of the compensating local Lorentz transformation that renders
$\delta_{\xi}T$ covariant turns out to be given by\footnote{After
  Ref.~\cite{Jacobson:2015uqa}, this parameter is often written in the
  equivalent, but less transparent, form
  \begin{equation}
    \label{eq:Jacobson}
  \sigma_{\xi}{}^{ab} =-\pounds_{\xi}e^{[a}{}_{\mu}e^{b]\, \mu}\, .  
  \end{equation}
}

\begin{equation}
  \label{eq:LLparameter}
  \sigma_{\xi}{}^{ab}
  =
  \imath_{\xi}\omega^{ab} -\nabla^{[a}\xi^{b]}\, ,
\end{equation}

\noindent
and it should be compared with the parameter of the compensating U(1) gauge
transformation $\chi_{\xi}$ in Eq.~(\ref{eq:LMparameter}). By analogy we can
define the Lorentz-algebra-valued \textit{momentum map}

\begin{equation}
  \label{eq:Pxiab}
  P_{\xi}{}^{ab} \equiv \nabla^{[a}\xi^{b]}\, .
\end{equation}

\noindent
We will see that this object satisfies a generalization of the equation that
defines the momentum map in the Maxwell case Eq.~(\ref{eq:Pkdef}).

It is natural to define the \textit{Lorentz-covariant Lie derivative} (or
\textit{Lie-Lorentz derivative}) of any tensor $T$ with Lorentz and world
indices with respect to a vector field $\xi$ as (minus) this
transformation:\footnote{The Lie-Lorentz derivative was originally introduced
  for spinor fields in Refs.~\cite{kn:Lich,kn:Kos,kn:Kos2,Hurley:cf} and its
  definition was later extended to more general Lorentz tensors $T$
  transforming in an arbitrary representation $r$ \cite{Ortin:2002qb}}

\begin{equation}
  \label{eq:LLderivative1}
  \mathbb{L}_{\xi}T
  \equiv
  -\delta_{\xi}T
  =
  \pounds_{\xi}T-\delta_{\sigma_{\xi}}T\, . 
\end{equation}

The properties of the Lie-Lorentz derivative on spinors are reviewed in
Refs.~\cite{Ortin:2002qb,Ortin:2015hya}. Here we are mainly interested in the
Lie-Lorentz derivatives of the Vielbein and the spin connection, specially
with respect to Killing vectors. According to the general definition, and
after trivial manipulations, we find that the Lie-Lorentz derivative of the
Vielbein is proportional to the Killing equation

\begin{equation}
  \label{eq:LLeam}
  \mathbb{L}_{\xi}e^{a}{}_{\mu}
  =
  \tfrac{1}{2}\left(\nabla_{\mu}\xi^{a}+\nabla^{a}\xi_{\mu}\right)
  =
  \tfrac{1}{2}e^{a\, \nu}\left(\nabla_{\mu}\xi_{\nu}+\nabla_{\nu}\xi_{\mu}\right)\, ,
\end{equation}

\noindent
and, therefore, it vanishes when $\xi$ is a Killing vector
field, independently of the basis chosen, as we should have expected.

We will use this equivalent differential-form expression for the above
equation:

\begin{equation}
  \label{eq:LLeam2}
  \mathbb{L}_{\xi}e^{a}
  =
  \mathcal{D}\xi^{a}+P_{\xi}{}^{a}{}_{b}e^{b}\, .
\end{equation}

Let us now consider the Lie-Lorentz derivative of the spin connection
$\omega^{ab}$. Taking into account the inhomogeneous form of the compensating
Lorentz transformation for the spin connection Eq.~(\ref{eq:deltasigmaomega})
we get\footnote{The same expression can be found if one considers the
  variation of the Levi-Civita spin connection as a function of the variation
  of the Vielbein, given by (minus) the Lie-Lorentz derivative in
  Eq.~(\ref{eq:LLeam}).}

\begin{equation}
  \label{eq:LLomega}
  \mathbb{L}_{\xi}\omega^{ab}
  =
  \pounds_{\xi}\omega^{ab}-\mathcal{D}\sigma_{\xi}{}^{ab}\, ,
\end{equation}

\noindent
where $\sigma_{\xi}{}^{ab}$ is with the same parameter
Eq.~(\ref{eq:LLparameter}). After some massaging, we can rewrite it in
a much more suggestive form

\begin{equation}
  \label{eq:LLomega2}
  \mathbb{L}_{\xi}\omega^{ab}
  =
  \imath_{\xi}R^{ab} +\mathcal{D}P_{\xi}{}^{ab}\, ,
\end{equation}

\noindent
where the Lorentz curvature 2-form
$R^{ab} \equiv \tfrac{1}{2}R_{\mu\nu}{}^{ab}dx^{\mu}\wedge dx^{\nu}$ is defined
as

\begin{equation}
\label{eq:Rmnab}
  R^{ab}
  =
  d\omega^{ab} -\omega^{a}{}_{c}\wedge \omega^{cb}\, ,
\end{equation}

\noindent
and where we have replaced $\nabla^{[a}\xi^{b]}$ by $P_{\xi}{}^{ab}$,
according to the definition of Eq.~(\ref{eq:Pxiab}).

The left-hand side of Eq.~(\ref{eq:LLomega2}) can be shown to vanish
identically when $\xi$ is a Killing vector field, because of the identity

\begin{equation}
  \label{eq:Pab1}
  \xi^{\nu}R_{\nu\mu}{}^{ab}+\nabla_{\mu}(\nabla^{[a}\xi^{b]})
  =
  \nabla^{[a}\left(\nabla^{b]}\xi_{\mu}+\nabla_{\mu}\xi^{b]}\right)\, .
\end{equation}

\noindent
As desired, for Killing vectors $k$ we have $\mathbb{L}_{k}e^{a}=0$ and
$\mathbb{L}_{k}\omega^{ab}=0$ and both statements are
Lorentz-invariant.\footnote{Observe that $\mathbb{L}_{\xi}\omega^{ab}$
  transforms as a Lorentz tensor even though $\omega^{ab}$ is not (it is a
  connection).}

For Killing vectors, Eq.~(\ref{eq:Pab1}) can also be written in the form 

\begin{equation}
\imath_{k}R^{ab} = -\mathcal{D}P_{k}{}^{ab}\, ,
\end{equation}

\noindent
which is the generalization of Eq.~(\ref{eq:Pkdef}) and justifies our definition
of momentum map Eq.~(\ref{eq:Pxiab}) for Killing vectors. The main difference
with the Lie-Maxwell case is that here we have an explicit expression for
$P_{\xi}{}^{ab}$ for any $\xi$.

In the context of asymptotically-flat stationary black holes, it is known
that, when evaluated on the event (Killing) horizon

\begin{equation}
  \label{eq:Pkab}
P_{k}{}^{ab}= \nabla^{[a}k^{b]} \stackrel{\mathcal{H}}{=} \kappa n^{ab}\, ,  
\end{equation}

\noindent
where $\kappa$ is the surface gravity and $n^{ab}$ is the binormal, normalized
to satisfy $n^{ab}n_{ab}=-2$. The constant\footnote{See
  Ref.~\cite{Wald:1984rg} for a proof of the constancy of $\kappa$ over the
  horizon (the standard zeroth law of black hole mechanics
  \cite{Bardeen:1973gs}) that makes use of the Einstein equations and the
  dominant energy condition and Ref.~\cite{Wald:1991tx} for a proof and does
  not, relaying only on the assumption of geodesic completeness of the null
  generators of the event horizon.} $\kappa$ is related to the Lorentz
momentum map just as the electric potential on the horizon was shown to be
related to the Maxwell momentum map in Section~\ref{sec-LMderivatives}. This
parallelism between zeroth laws was observed in \cite{Prabhu:2015vua}.

%%%%%%%%%%%%%%%%%%%%%%%%%%%%%%%%%%%%%%%%%%%%%%%%%%%%%%%%%%%%%%%%%%%%%%
%%%%%%%%%%%%%%%%%%%%%%%%%%%%%%%%%%%%%%%%%%%%%%%%%%%%%%%%%%%%%%%%%%%%%%
%%%%%%%%%%%%%%%%%%%%%%%%%%%%%%%%%%%%%%%%%%%%%%%%%%%%%%%%%%%%%%%%%%%%%%
%%%%%%%%%%%%%%%%%%%%%%%%%%%%%%%%%%%%%%%%%%%%%%%%%%%%%%%%%%%%%%%%%%%%%%
\section{The Einstein-Maxwell action and the RNT solutions}
\label{sec-setup} 
%%%%%%%%%%%%%%%%%%%%%%%%%%%%%%%%%%%%%%%%%%%%%%%%%%%%%%%%%%%%%%%%%%%%%%
%%%%%%%%%%%%%%%%%%%%%%%%%%%%%%%%%%%%%%%%%%%%%%%%%%%%%%%%%%%%%%%%%%%%%%
%%%%%%%%%%%%%%%%%%%%%%%%%%%%%%%%%%%%%%%%%%%%%%%%%%%%%%%%%%%%%%%%%%%%%%
%%%%%%%%%%%%%%%%%%%%%%%%%%%%%%%%%%%%%%%%%%%%%%%%%%%%%%%%%%%%%%%%%%%%%%

In this section we present the $d$-dimensional Einstein theory and the
$d$-dimensional Reissner-Nordstr\"om-Tangherlini (RNT) solutions we are going
to study, in order to fix the conventions. We will first give the action and
equations of motion in the standard tensorial form, and will then rewrite
them in the differential-language form that we will use in the following section.

%%%%%%%%%%%%%%%%%%%%%%%%%%%%%%%%%%%%%%%%%%%%%%%%%%%%%%%%%%%%%%%%%%%%%%
%%%%%%%%%%%%%%%%%%%%%%%%%%%%%%%%%%%%%%%%%%%%%%%%%%%%%%%%%%%%%%%%%%%%%%
%%%%%%%%%%%%%%%%%%%%%%%%%%%%%%%%%%%%%%%%%%%%%%%%%%%%%%%%%%%%%%%%%%%%%%
%%%%%%%%%%%%%%%%%%%%%%%%%%%%%%%%%%%%%%%%%%%%%%%%%%%%%%%%%%%%%%%%%%%%%%
\subsection{Action and equations of motion}
\label{sec-actionandeoms}
%%%%%%%%%%%%%%%%%%%%%%%%%%%%%%%%%%%%%%%%%%%%%%%%%%%%%%%%%%%%%%%%%%%%%%
%%%%%%%%%%%%%%%%%%%%%%%%%%%%%%%%%%%%%%%%%%%%%%%%%%%%%%%%%%%%%%%%%%%%%%
%%%%%%%%%%%%%%%%%%%%%%%%%%%%%%%%%%%%%%%%%%%%%%%%%%%%%%%%%%%%%%%%%%%%%%
%%%%%%%%%%%%%%%%%%%%%%%%%%%%%%%%%%%%%%%%%%%%%%%%%%%%%%%%%%%%%%%%%%%%%%

Setting $G_{N}^{(d)}=1$ for simplicity, and choosing as basic dynamical
fields the Vielbein $e^{a}{}_{\mu}$ and the Maxwell field $A_{\mu}$, the
action of the Einstein-Maxwell theory in $d$ spacetime dimensions

\begin{equation}
\label{eq:EMaction}
S[e^{a}{}_{\mu},A_{\mu}]
=
\frac{1}{16\pi}\int d^{d}x\, e\, \left[R(\omega,e) -\tfrac{1}{4}F^{2}\right]\, .
\end{equation}

\noindent
where $e\equiv \text{det}(e^{a}{}_{\mu})$, $R(\omega,e)$ is the Ricci scalar,
defined in terms of the Levi-Civita spin connection
$\omega_{\mu}{}^{ab}$,\footnote{We are using the second-order formalism.}
that is

\begin{equation}
R(\omega,e) = e_{a}{}^{\mu}e_{b}{}^{\nu}R_{\mu\nu}{}^{ab}(\omega)\, ,  
\end{equation}

\noindent
where $R_{\mu\nu}{}^{ab}(\omega)$ is the curvature 2-form of the Levi-Civita
spin connection, defined in Eq.~(\ref{eq:Rmnab}). The Levi-Civita spin
connection (metric compatible and torsion-free, that is $\mathcal{D}e^{a}=0$)
is given by

\begin{equation}
  \omega_{abc} = e_{a}{}^{\mu}\omega_{\mu\, ba} = -\Omega_{abc}+\Omega_{bca}-\Omega_{cab}\, ,
  \hspace{1cm}
  \Omega_{abc}= e_{a}{}^{\mu}e_{b}{}^{\nu} \partial_{[\mu|}e_{c\, |\nu]}\, .
\end{equation}

Finally, $F^{2}=F_{ab}F^{ab}$, $F_{ab} = e_{a}{}^{\mu}e_{b}{}^{\nu}F_{\mu\nu}$
and $F_{\mu\nu}$ is defined in Eq.~(\ref{eq:Fmn}).

The equations of motion are

\begin{subequations}
  \begin{align}
    \label{eq:Eam}
    E_{a}{}^{\mu}
    & \equiv
      \frac{\delta S}{\delta e^{a}{}_{\mu}}
    =
    -\frac{e}{8\pi}\left(G_{a}{}^{\mu} -\tfrac{1}{2}T^{a}{}_{\mu}\right)\, ,
    \\
    & \nonumber \\
    \label{eq:Em}
    E^{\mu}
    & \equiv
      \frac{\delta S}{\delta A_{\mu}}
     =
    \frac{1}{16\pi}\partial_{\nu}\left(eF^{\nu\mu}\right)\, ,
  \end{align}
\end{subequations}

\noindent
where

\begin{equation}
  \label{eq:Tam}
  T_{a}{}^{\mu}
  =
F_{ab}F^{\mu b} -\tfrac{1}{4}e_{a}{}^{\mu}F^{2}\, ,
\end{equation}

\noindent
is the electromagnetic field's energy-momentum tensor.

In differential-form language, the action Eq.~(\ref{eq:EMaction}) is usually
written in this form

\begin{equation}
\label{eq:EMaction2}
S[e^{a},A]
=
\frac{(-1)^{d-1}}{16\pi}\int 
\left[
  \frac{1}{(d-2)!}R^{a_{1}a_{2}}\wedge e^{a_{3}}\wedge \cdots \wedge
  e^{a_{d}}\epsilon_{a_{1}\cdots a_{d}}
  -\tfrac{1}{2}F\wedge \star F
\right]
\equiv
\int \mathbf{L}\, ,
\end{equation}

\noindent
although it is more convenient to rewrite the first (Einstein-Hilbert)  term
as

\begin{equation}
  \frac{1}{(d-2)!}R^{a_{1}a_{2}}\wedge e^{a_{3}}\wedge \cdots \wedge
  e^{a_{d}}\epsilon_{a_{1}\cdots a_{d}}
  =
  \star (e^{a}\wedge e^{b}) \wedge R_{ab}\, .
\end{equation}

The $(d-1)$-form equations of motion (which we write in boldface) are
given by

\begin{subequations}
  \begin{align}
    \label{eq:Eb}
  \mathbf{E}_{a}
  & =
    \frac{1}{16\pi}
    \left\{
    \imath_{a}\star (e^{c}\wedge e^{d})\wedge R_{cd}
    +\tfrac{1}{2}\left(\imath_{a}F\wedge \star F-F\wedge \imath_{a}\star F\right)
    \right\}\, ,
\\
    & \nonumber \\
    \label{eq:E}
    \mathbf{E}
    & =
      -\frac{1}{16\pi}d\star F\, ,
  \end{align}
\end{subequations}

\noindent
where $\imath_{c}$ stands for $i_{e_{c}}$, where
$e_{c}=e_{c}{}^{\mu}\partial_{\mu}$.

%%%%%%%%%%%%%%%%%%%%%%%%%%%%%%%%%%%%%%%%%%%%%%%%%%%%%%%%%%%%%%%%%%%%%%
%%%%%%%%%%%%%%%%%%%%%%%%%%%%%%%%%%%%%%%%%%%%%%%%%%%%%%%%%%%%%%%%%%%%%%
%%%%%%%%%%%%%%%%%%%%%%%%%%%%%%%%%%%%%%%%%%%%%%%%%%%%%%%%%%%%%%%%%%%%%%
%%%%%%%%%%%%%%%%%%%%%%%%%%%%%%%%%%%%%%%%%%%%%%%%%%%%%%%%%%%%%%%%%%%%%%
\subsection{The Reissner-Nordstr\"om-Tangherlini solutions}
\label{sec-RTNsolutions}
%%%%%%%%%%%%%%%%%%%%%%%%%%%%%%%%%%%%%%%%%%%%%%%%%%%%%%%%%%%%%%%%%%%%%%
%%%%%%%%%%%%%%%%%%%%%%%%%%%%%%%%%%%%%%%%%%%%%%%%%%%%%%%%%%%%%%%%%%%%%%
%%%%%%%%%%%%%%%%%%%%%%%%%%%%%%%%%%%%%%%%%%%%%%%%%%%%%%%%%%%%%%%%%%%%%%
%%%%%%%%%%%%%%%%%%%%%%%%%%%%%%%%%%%%%%%%%%%%%%%%%%%%%%%%%%%%%%%%%%%%%%

The $d$-dimensional RNT solutions with rationalized mass $M$ and electric
charge $q$ are described by the following metric and electromagnetic fields
\cite{kn:Reiss,kn:No8,Tangherlini:1963bw}:

\begin{equation}
  \label{eq:RNTsolution}
    ds^{2}
     =
    \lambda dt^{2} - \frac{dr^{2}}{\lambda} -r^{2}d\Omega^{2}_{(d-2)}\, ,
          \hspace{1cm}
    F_{tr}
    =
      \frac{16\pi}{\omega_{(d-2)}}\frac{q}{r^{d-2}}\, ,
\end{equation}

\noindent
where $d\Omega^{2}_{(d-2)}$ is the metric of the round $(d-2)$-sphere of unit
radius, $\omega_{(d-2)}$ is its volume and  
 
\begin{subequations}
  \begin{align}
    \lambda
    & =
      \frac{(r^{d-3}-r^{d-3}_{+})(r^{d-3}-r^{d-3}_{-})}{r^{2(d-3)}}\, ,
    \\
    & \nonumber \\
    r^{d-3}_{\pm}
    & =
      \frac{8\pi}{(d-2)\omega_{(d-2)}} M \pm r^{d-3}_{0}\, ,
    \\
    & \nonumber \\
    r^{d-3}_{0}
    & =
    \frac{8\pi}{(d-2)\omega_{(d-2)}} \sqrt{M^{2} -\frac{2(d-2)}{(d-3)}q^{2}}\, .
  \end{align}
\end{subequations}

\noindent
The origin of the annoying normalization factors lies in the standard
normalization factor $(16\pi)^{-1}$ of the action, which should be replaced by
$[2(d-2)\omega_{(d-2)}]^{-1}$. Instead, we can just define

\begin{equation}
  \mathcal{M} \equiv \frac{8\pi}{(d-2)\omega_{(d-2)}} M\, ,
  \hspace{1cm}
  \mathcal{Q} \equiv \frac{16\pi}{\omega_{(d-2)}} q\, ,
\end{equation}

\noindent
getting somewhat simpler expressions

\begin{subequations}
  \begin{align}
 F_{tr}
    & =
    \frac{\mathcal{Q}}{r^{d-2}}\, ,
    \\
    & \nonumber \\
    r^{d-3}_{\pm}
    & =
      \mathcal{M} \pm r^{d-3}_{0}\, ,
    \\
    & \nonumber \\
    r^{d-3}_{0}
    & =
    \sqrt{\mathcal{M}^{2} -\frac{\mathcal{Q}^{2}}{2(d-2)(d-3)}}\, .
  \end{align}
\end{subequations}

The event horizon of these solutions exists when
$\mathcal{M} \geq [2(d-2)(d-3)]^{-1/2}|\mathcal{Q}|$ and then it is located at
$r=r_{+}$ and its surface gravity is given by

\begin{equation}
  \kappa=(d-3) r^{d-3}_{0}/r^{d-2}_{+}\, .
\end{equation}

\noindent
The surface gravity vanishes in the extremal limit $r_{0}=0$, which is reached
when $\mathcal{M} = [2(d-2)(d-3)]^{-1/2}|\mathcal{Q}|$. We will always assume
that $\kappa\neq 0$.

The timelike Killing vector that becomes null on the horizon is
$k=\partial_{t}$ in these coordinates, but they do not cover the bifurcate
sphere because this expression for $k$ never vanishes. In the region covered
by these coordinates we find that

\begin{equation}
  P_{k}{}^{\mu\nu}
  =
  \nabla^{[\mu}k^{\nu]}
  =
  -\partial_{r}\lambda g^{\mu\nu}{}_{rt}
  \stackrel{\mathcal{H}}{=}  \kappa n^{\mu\nu}\, ,
\end{equation}

\noindent
where the binormal takes the value

\begin{equation}
  n^{\mu\nu} = -2 g^{\mu\nu}{}_{rt} \, ,
  \,\,\,\,\,
  \Rightarrow
  \,\,\,\,\,
  n^{\mu\nu}n_{\mu\nu} = -2\, .
\end{equation}

On the other hand, $\imath_{k}F=F_{tr}dr$ and

\begin{equation}
  \label{eq:PhiRNT}
  P_{k}
  =
  \frac{\mathcal{Q}/(d-3)}{r^{d-3}}
  \stackrel{\mathcal{H}}{=}
  \frac{\mathcal{Q}/(d-3)}{r_{+}^{d-3}}
  =
  \Phi\, .
\end{equation}

In order to reach the bifurcation sphere we need to use Kruskal-Szekeres
coordinates.  For $d=4$ the change from $r,t$ to Kruskal-Szekeres's $U,V$ is
known and given explicitly, for instance, in Ref.~\cite{Hawking:1973uf}. To
work in arbitrary $d$ we will just work near the event horizon: expanding the
solution in Eq.~(\ref{eq:RNTsolution}) around $r=r_{+}$ and ignoring terms of
second or higher order in $r-r_{+}$ we get

\begin{subequations}
  \label{eq:RNTsolutionNH}
  \begin{align}
    ds^{2}
     & =
     2\kappa (r-r_{+}) dt^{2} - \frac{dr^{2}}{ 2\kappa (r-r_{+})}
     -r_{+}^{2}\left[1+2(r-r_{+})/r_{+}\right]d\Omega^{2}_{(d-2)}
     +\mathcal{O}(r-r_{+})^{2}\, ,
    \\
    & \nonumber \\
    F_{tr}
     &  =
      \frac{\mathcal{Q}}{r_{+}^{d-2}}\left[1 -(d-2)(r-r_{+})/r_{+}\right]
       +\mathcal{O}(r-r_{+})^{2}\, .
  \end{align}
\end{subequations}

The tortoise coordinate $r_{*}$ is

\begin{equation}
  r_{*} = \frac{1}{2\kappa}\log{\left(\frac{r-r_{+}}{r_{+}}\right)}
  +C +\mathcal{O}(r-r_{+})^{2}\, ,
\end{equation}

\noindent
where $C$ is an integration constant that we set to zero for the sake of
convenience. Defining

\begin{equation}
  v\equiv t+r_{*}\, ,
  \hspace{1cm}
  u\equiv t-r_{*}\, ,
\end{equation}

\noindent
the solution takes the form

\begin{subequations}
  \label{eq:RNTsolutionNH2}
  \begin{align}
    ds^{2}
     & =
     2\kappa r_{+} e^{\kappa(v-u)} dudv
     -r_{+}^{2}\left[1+2 e^{\kappa(v-u)}\right]d\Omega^{2}_{(d-2)}
     +\mathcal{O}(r-r_{+})^{2}\, ,
    \\
    & \nonumber \\
    F_{uv}
    &  =
      \kappa \frac{\mathcal{Q}}{r_{+}^{d-3}}  e^{\kappa(v-u)}
       +\mathcal{O}(r-r_{+})^{2}\, .
  \end{align}
\end{subequations}

Finally, we define the coordinates $U,V$

\begin{equation}
  V \equiv \sqrt{r_{+}/\kappa}\, e^{\kappa v}\, ,
  \hspace{1cm}
  U \equiv - \sqrt{r_{+}/\kappa}\, e^{-\kappa u}\, ,
\end{equation}

\noindent
in terms of which the solution takes the form

\begin{subequations}
  \label{eq:RNTsolutionNH3}
  \begin{align}
    ds^{2}
     & =
     -2dUdV
     -r_{+}^{2}\left[1-2\kappa UV/r_{+}\right]d\Omega^{2}_{(d-2)}
     +\mathcal{O}(UV)^{2}\, ,
    \\
    & \nonumber \\
    F_{UV}
    &  =
      -\frac{\mathcal{Q}}{r_{+}^{d-2}}
       +\mathcal{O}(UV)^{2}\, .
  \end{align}
\end{subequations}

The Killing vector $k=\partial_{t}$ becomes, in these coordinates

\begin{equation}
  k=\kappa\left(V\partial_{V}-U\partial_{U}\right) +\mathcal{O}(UV)^{2}\, ,
  \hspace{1cm}
  \hat{k}\equiv k_{\mu}dx^{\mu} = \kappa\left( VdU-UdV\right)+\mathcal{O}(UV)^{2}\, .
\end{equation}

In these coordinates, the hypersurface $U=0$ is the past event horizon
$\mathcal{H}^{-}$, generated by
$\left. k\right|_{\mathcal{H}^{-}} = \kappa V\partial_{V}= \partial_{v}$.  The
hypersurface $V=0$ is the future event horizon $\mathcal{H}^{+}$ generated by
$\left. k\right|_{\mathcal{H}^{+}} = -\kappa U\partial_{U}=
\partial_{u}$. They cross at the bifurcation sphere, which is defined by
$U=V=0$ and can also be characterized as the spatial cross section of the
horizon at which $k=0$.

On the other hand,

\begin{equation}
  \begin{aligned}
    P_{k\, \mu\nu}dx^{\mu}\wedge dx^{\nu}
    & =
    d\hat{k} = 2\kappa dV\wedge dU
    +\mathcal{O}(UV)^{2} = 2\kappa g_{VU, \mu\nu}dx^{\mu}\wedge dx^{\nu}
    +\mathcal{O}(UV)^{2}\, ,
    \\
    & \\
    \Rightarrow \,\,\,\,\, n_{\mu\nu}
    & =
    -2g_{UV,\, \mu\nu}\, .
  \end{aligned}
\end{equation}

On the other hand,

\begin{equation}
  \begin{aligned}
    \imath_{k}F & = \kappa \frac{\mathcal{Q}}{r_{+}^{d-2}} (VdU+UdV)
    +\mathcal{O}(UV)^{2}\, ,
    \\
    & \\
    \Rightarrow \,\,\,\,\, P_{k}
    & =
    C +\kappa
    \frac{\mathcal{Q}}{r_{+}^{d-2}}UV +\mathcal{O}(UV)^{2}\, .
  \end{aligned}
\end{equation}

\noindent
The constant $C$ clearly has to be identified with the electric potential
over the horizon $\Phi$ in Eq.~(\ref{eq:PhiRNT}).  As observed in
Ref.~\cite{Gao:2003ys}, if we use the simplest choice of electromagnetic
potential

\begin{equation}
A =   \frac{\mathcal{Q}/(d-3)}{r^{d-3}}dt\, ,
\end{equation}

\noindent
we obtain,

\begin{equation}
  A = \frac{\mathcal{Q}}{2(d-3)\kappa r_{+}^{d-3}}
  \left[1 +(d-3)\kappa UV/r_{+}  +\mathcal{O}(UV)^{2}\right]
  \left(\frac{dV}{V}-\frac{dU}{U}\right)\, ,
\end{equation}

\noindent
which is singular at the horizon.

%%%%%%%%%%%%%%%%%%%%%%%%%%%%%%%%%%%%%%%%%%%%%%%%%%%%%%%%%%%%%%%%%%%%%% 
%%%%%%%%%%%%%%%%%%%%%%%%%%%%%%%%%%%%%%%%%%%%%%%%%%%%%%%%%%%%%%%%%%%%%%
%%%%%%%%%%%%%%%%%%%%%%%%%%%%%%%%%%%%%%%%%%%%%%%%%%%%%%%%%%%%%%%%%%%%%%
%%%%%%%%%%%%%%%%%%%%%%%%%%%%%%%%%%%%%%%%%%%%%%%%%%%%%%%%%%%%%%%%%%%%%%
\section{Wald-Noether charge for the E-M theory}
\label{sec-WNcharge}
%%%%%%%%%%%%%%%%%%%%%%%%%%%%%%%%%%%%%%%%%%%%%%%%%%%%%%%%%%%%%%%%%%%%%%
%%%%%%%%%%%%%%%%%%%%%%%%%%%%%%%%%%%%%%%%%%%%%%%%%%%%%%%%%%%%%%%%%%%%%%
%%%%%%%%%%%%%%%%%%%%%%%%%%%%%%%%%%%%%%%%%%%%%%%%%%%%%%%%%%%%%%%%%%%%%%
%%%%%%%%%%%%%%%%%%%%%%%%%%%%%%%%%%%%%%%%%%%%%%%%%%%%%%%%%%%%%%%%%%%%%%

The general variation of the action of the Einstein-Maxwell theory
Eq.~(\ref{eq:EMaction2}) is

\begin{equation}
    \delta S
     =
     \int 
     \left\{
       \mathbf{E}_{a}\wedge \delta e^{a} +\mathbf{E}\wedge \delta A
       +d\mathbf{\Theta}(e,A,\delta e,\delta A)
    \right\}\, ,
\end{equation}

\noindent
where $\mathbf{E}_{a}$ and $\mathbf{E}$ are, respectively, the $(d-1)$-form
Einstein (\ref{eq:Eb}) and Maxwell (\ref{eq:Em}) equations multiplied by the
volume form $d^{d}x$ and

\begin{equation}
  \mathbf{\Theta}(e,A,\delta e,\delta A)
  \equiv
  -\frac{1}{16\pi}\left[
    \star (e^{a}\wedge e^{b})\wedge \delta \omega_{ab}
    -\star F \wedge \delta A\right]\, ,
%   \frac{\epsilon_{\alpha_{1}\cdots \alpha_{d-1}\mu}}{(d-1)!\, e}
%   \left(2 e_{a}{}^{\mu} e_{b}{}^{\nu}\delta \omega_{\nu}{}^{ab}(e)
%     -F^{\mu\nu}\delta A_{\nu}\right)
%   dx^{\alpha_{1}}\wedge \cdots \wedge dx^{\alpha_{d-1}}\, ,
\end{equation}

\noindent
is the \textit{presymplectic $(d-1)$-form} defined in Ref.~\cite{Lee:1990nz}
and $\star$ stands for the Hodge dual. For the transformations given by
(minus) the covariant Lie derivatives in Eqs.~(\ref{eq:LieMaxwelldef}),
(\ref{eq:LLeam2}) and (\ref{eq:LLomega2})

\begin{equation}
  \label{eq:dSxi}
    \delta_{\xi} S
     =
     \int 
     \left\{
       -\mathbf{E}_{a}\wedge (\mathcal{D}\xi^{a} +P_{\xi}{}^{a}{}_{b}e^{b})
       -\mathbf{E}\wedge (\imath_{\xi}F +dP_{\xi})
       +d\mathbf{\Theta}(e,A,\delta_{\xi} e,\delta_{\xi} A)
    \right\}\, ,
\end{equation}

\noindent
with

\begin{equation}
  \mathbf{\Theta}(e,A,\delta_{\xi} e,\delta_{\xi} A)
  =
 \frac{1}{16\pi}\left[ \star ( e^{a}\wedge e^{b})\wedge
    \left(\imath_{\xi}R^{ab} +\mathcal{D}P_{\xi}{}^{ab} \right)
    -\star F \wedge  \left(\imath_{\xi}F +dP_{\xi}\right) \right]\, .
%   \frac{\epsilon_{\alpha_{1}\cdots \alpha_{d-1}\mu}}{(d-1)!\, e}
%   \left(2 e_{a}{}^{\mu} e_{b}{}^{\nu}\delta \omega_{\nu}{}^{ab}(e)
%     -F^{\mu\nu}\delta A_{\nu}\right)
%   dx^{\alpha_{1}}\wedge \cdots \wedge dx^{\alpha_{d-1}}\, ,
\end{equation}

Let us consider the first term. It is not difficult to see that
$\mathbf{E}_{a}\wedge e^{b} P_{\xi}{}^{a}{}_{b}=0$ because the tensor
contracted with the Lorentz momentum map give the Einstein equations, which
are symmetric in the indices $a$ and $b$. The rest can be integrated by parts,

\begin{equation}
  -\mathbf{E}_{a}\wedge \mathcal{D}\xi^{a}
   =
  -(-1)^{d-1}d\left(\mathbf{E}_{a}\xi^{a}\right)
    +(-1)^{d-1}\xi^{a}\mathcal{D}\mathbf{E}_{a}\, .
\end{equation}

\noindent
Using the Bianchi identity $\mathcal{D}R^{ab}=0$, 

\begin{equation}
  \begin{aligned}
    \xi^{a}\mathcal{D}\mathbf{E}_{a}
    & =
    \frac{1}{32\pi}\xi^{a}
    \mathcal{D}\left(\imath_{a}F\wedge \star F-F\wedge \imath_{a}\star F\right)
    \\
    & \\
    & =
    \frac{1}{32\pi}\xi^{a}
    \left[\mathbf{\nabla}\imath_{a}F\wedge \star F -\imath_{a}F\wedge \mathbf{\nabla}\star F
      -\mathbf{\nabla}F\wedge \imath_{a}\star F
      -F\wedge \mathbf{\nabla}\imath_{a}\star F\right]\, ,
  \end{aligned}
\end{equation}

\noindent
where we have replaced $\mathcal{D}$ by $\mathbf{\nabla}$ is the exterior
total covariant derivative operator which satisfies the first Vielbein
postulate. Then, using the property

\begin{equation}
 \mathbf{\nabla}\imath_{a}\omega = -\imath_{a}d\omega +\nabla_{a}\omega\, ,
\end{equation}

\noindent
and replacing $\mathbf{\nabla}$ by the exterior derivative when it acts on
differential forms with no indices, as well as using the Bianchi identity
$dF=0$, we get

\begin{equation}
    \xi^{a}\mathcal{D}\mathbf{E}_{a}
    =
    \frac{1}{32\pi}\xi^{a}
    \left[
%      -\imath_{a}dF\wedge \star F +
      \nabla_{a}F\wedge \star F
      -\imath_{a}F\wedge d\star F
 %     -dF\wedge \imath_{a}\star F
      +F\wedge \imath_{a}d\star F
      -F\wedge \nabla_{a}\star F
    \right]\, .
\end{equation}

\noindent
Since $\nabla_{a}$ commutes with the Hodge dual and
$F \wedge \star G$ is symmetric in $F$ and $G$ for any 2-forms $F,G$, the two
terms with $\nabla_{a}$ cancel each other. Furthermore,

\begin{equation}
  F\wedge \imath_{a}d\star F \
  =
\imath_{a}(F\wedge d\star F) - \imath_{a}F\wedge d\star F\, ,
\end{equation}

\noindent
and

\begin{equation}
\xi^{a}\imath_{a}\omega = \imath_{\xi}\omega\, ,  
\end{equation}

\noindent
for any $p$-form, we arrive at

\begin{equation}
(-1)^{d-1}\xi^{a}\mathcal{D}\mathbf{E}_{a}
    =
-\frac{1}{16\pi} d\star F\wedge \imath_{\xi}F\, .
\end{equation}

The second term in Eq.~(\ref{eq:dSxi}) gives

\begin{equation}
    -\mathbf{E}\wedge (\imath_{\xi}F +dP_{\xi})
    =
    \frac{1}{16\pi}d\star F\wedge \imath_{\xi}F
    -(-1)^{d-1}d\left(\mathbf{E}P_{\xi}\right)\, ,
\end{equation}

\noindent
and, collecting the partial results, we get

\begin{equation}
  \label{eq:dSxi2}
    \delta S_{\xi}
     =
     \int  d\mathbf{\Theta}'(e,A,\delta_{\xi} e,\delta_{\xi} A)\, ,
\end{equation}

\noindent
where

\begin{equation}
  \label{eq:Thetaprime}
  \begin{aligned}
    \mathbf{\Theta}'(e,A,\delta_{\xi} e,\delta_{\xi} A)
    & \equiv
    \mathbf{\Theta}(e,A,\delta_{\xi} e,\delta_{\xi} A)
    +(-1)^{d}\left(\mathbf{E}_{a}\xi^{a}+\mathbf{E}P_{\xi}\right)
    \\
    & \\
    & =
    \frac{1}{16\pi}\left[ \star ( e^{a}\wedge e^{b})\wedge
    \left(\imath_{\xi}R_{ab} +\mathcal{D}P_{\xi\, ab} \right)
    -\star F \wedge  \left(\imath_{\xi}F +dP_{\xi}\right) \right.
  \\
  & \\
  &
  +(-1)^{d}\imath_{\xi}\star (e^{a}\wedge e^{b})\wedge R_{ab}
    +\frac{(-1)^{d}}{2}\left(\imath_{\xi}F\wedge \star F-F\wedge \imath_{\xi}\star F\right)
    \\
    & \\
    &
    \left.
+(-1)^{d-1}d\star F P_{\xi}
  \right]
  \\
  & \\
  & =
  -\imath_{\xi}\mathbf{L}
  +\frac{(-1)^{d-1}}{16\pi}d\left[\star F P_{\xi}
    -\star (e^{a}\wedge e^{b})P_{\xi\, ab}\right]\, .
\end{aligned}
\end{equation}

The action of the Einstein-Maxwell theory Eq.~(\ref{eq:EMaction2}) is exactly
invariant under local Lorentz and electromagnetic gauge transformations and it
is invariant up to a total derivative under diffeomorphisms. Therefore, under
the combined transformations $\delta_{\xi} \equiv -\mathbb{L}_{\xi}$ with the
covariant Lie derivatives defined in Eqs.~(\ref{eq:LieMaxwelldef}),
(\ref{eq:LLeam2}) and (\ref{eq:LLomega2}),

\begin{equation}
\delta_{\xi}S = -\int d \imath_{\xi}\mathbf{L}\, .  
\end{equation}

\noindent
Taking into account the result in Eq.~(\ref{eq:dSxi2}), the arbitrariness of
the domain of integration, of the parameter $\xi$, and the fact that we have
not used the equations of motion, we conclude that, if we define the
$(d-1)$-form

\begin{equation}
  \mathbf{J}
  \equiv
  \mathbf{\Theta}'(e,A,\delta_{\xi} e,\delta_{\xi} A)+\imath_{\xi}\mathbf{L}\, ,
\end{equation}

\noindent
it satisfies 

\begin{equation}
d\mathbf{J} = 0\, ,
\end{equation}

\noindent
identically, off-shell. This, in its turn, implies the existence of a
$(d-2)$-form $\mathbf{Q}[\xi]$ (the Wald-Noether charge) such that

\begin{equation}
  \label{eq:J}
\mathbf{J} = d\mathbf{Q}[\xi]\, .
\end{equation}

\noindent
The last line of Eq.~(\ref{eq:Thetaprime}) gives the following expression for
the Wald-Noether charge:

\begin{equation}
  \label{eq:Q}
    \mathbf{Q}[\xi]
    =
      \frac{(-1)^{d-1}}{16\pi}\left[\star F P_{\xi}
    -\star (e^{a}\wedge e^{b}) P_{\xi\, ab}\right]\, .
\end{equation}

%%%%%%%%%%%%%%%%%%%%%%%%%%%%%%%%%%%%%%%%%%%%%%%%%%%%%%%%%%%%%%%%%%%%%%
%%%%%%%%%%%%%%%%%%%%%%%%%%%%%%%%%%%%%%%%%%%%%%%%%%%%%%%%%%%%%%%%%%%%%%
%%%%%%%%%%%%%%%%%%%%%%%%%%%%%%%%%%%%%%%%%%%%%%%%%%%%%%%%%%%%%%%%%%%%%%
%%%%%%%%%%%%%%%%%%%%%%%%%%%%%%%%%%%%%%%%%%%%%%%%%%%%%%%%%%%%%%%%%%%%%%
\section{The first law of black hole mechanics in the E-M theory}
\label{sec-1stlaw}
%%%%%%%%%%%%%%%%%%%%%%%%%%%%%%%%%%%%%%%%%%%%%%%%%%%%%%%%%%%%%%%%%%%%%%
%%%%%%%%%%%%%%%%%%%%%%%%%%%%%%%%%%%%%%%%%%%%%%%%%%%%%%%%%%%%%%%%%%%%%%
%%%%%%%%%%%%%%%%%%%%%%%%%%%%%%%%%%%%%%%%%%%%%%%%%%%%%%%%%%%%%%%%%%%%%%
%%%%%%%%%%%%%%%%%%%%%%%%%%%%%%%%%%%%%%%%%%%%%%%%%%%%%%%%%%%%%%%%%%%%%%

Following Ref.~\cite{Lee:1990nz} we define the pre-symplectic $(d-1)$-form

\begin{equation}
\omega(\phi,\delta_{1}\phi,\delta_{2} \phi)  
\equiv 
\delta_{1}\mathbf{\Theta}(\phi,\delta_{2} \phi)
-\delta_{2}\mathbf{\Theta}(\phi,\delta_{1} \phi)\, ,
\end{equation}

\noindent
where $\phi$ stands for the Vielbein and Maxwell fields, and the symplectic
form relative to the Cauchy surface $\Sigma$

\begin{equation}
\Omega(\phi,\delta_{1}\phi,\delta_{2} \phi)  
\equiv
\int_{\Sigma}\omega(\phi,\delta_{1}\phi,\delta_{2} \phi)\, .
\end{equation}

Following now Ref.~\cite{Iyer:1994ys}, when $\phi$ solves the equations of
motion $\mathbf{E}_{\phi}=0$, for any variation of the fields
$\delta_{1}\phi=\delta\phi$ and the variations under diffeomorphisms
$\delta_{2}\phi= \delta_{\xi}\phi$

\begin{equation}
  \omega(\phi,\delta\phi,\delta_{\xi}\phi)
  =
  \delta\mathbf{J}+d\imath_{\xi}\mathbf{\Theta}'
  =
  \delta d\mathbf{Q}[\xi]+d\imath_{\xi}\mathbf{\Theta}'\, ,
\end{equation}

\noindent
where, in our case, $\mathbf{J}$ is given by Eq.~(\ref{eq:J}),
$\mathbf{\Theta}'$ is given in Eq.~(\ref{eq:Thetaprime}) and we observe that,
on-shell, $\mathbf{\Theta} = \mathbf{\Theta}'$.  Then, if $\delta\phi$
satisfies the linearized equations of motion
$\delta d\mathbf{Q}= d\delta \mathbf{Q}$. Furthermore, if the parameter $\xi=k$
generates a transformation that leaves invariant all the fields of the theory,
$\delta_{k}\phi=0$, $\omega(\phi,\delta\phi,\delta_{k}\phi)=0$, and we
arrive at

\begin{equation}
d\left( \delta \mathbf{Q}[k]+\imath_{k}\mathbf{\Theta}'  \right)=0\, ,
\end{equation}

\noindent
which, when integrated over a hypersurface $\Sigma$ with boundary
$\delta\Sigma$, gives

\begin{equation}
  \int_{\delta\Sigma}  \left( \delta \mathbf{Q}[k]+\imath_{k}\mathbf{\Theta}'  \right)
  =
  0\, .
\end{equation}
 
In our case, we are dealing with asymptotically flat, static black holes. $k$
is the timelike Killing vector whose Killing horizon coincides with the event
horizon and the hypersurface $\Sigma$ is the space between infinity and the
bifurcation sphere ($\mathcal{BH}$) on which $k=0$. Infinity and the bifurcate
horizon are the two disconnected components of $\delta\Sigma$ and taking into
account that $k=0$ on the bifurcation sphere, we obtain

\begin{equation}
\delta \int_{\mathcal{BH}}   \mathbf{Q}[k]
=
\int_{\infty}  \left( \delta \mathbf{Q}[k]+\imath_{k}\mathbf{\Theta}'  \right)\, .
\end{equation}

As explained in Ref.~\cite{Iyer:1994ys}, the right-hand side can be identified
with $\delta M$, where $M$ is the total mass of the black-hole
spacetime. Using Eq.~(\ref{eq:Q}), we find

\begin{equation}
  \delta \int_{\mathcal{BH}}   \mathbf{Q}[k]
  =
  \frac{(-1)^{d-1}}{16\pi}\delta \int_{\mathcal{BH}} \star F P_{k}
  +\frac{(-1)^{d}}{16\pi} \delta \int_{\mathcal{BH}} \star
  (e^{a}\wedge e^{b})P_{k\, ab}\, .
\end{equation}

\noindent
According to the discussion at the end of Section~\ref{sec-LMderivatives},
$P_{k}$ can be identified with the electric potential $\Phi$ and it is
constant over the horizon. The electric charge contained inside the horizon 
is given by

\begin{equation}
  \mathcal{Q}
  \equiv
  \frac{(-1)^{d-1}}{16\pi}\int_{\mathcal{BH}} \star F\, ,
\end{equation}

\noindent
and the first term just gives $+\Phi \delta \mathcal{Q}$, which implies that
we get a first-law-like relation if the second term gives $T\delta S$. Let us
study that term. Using Eq.~(\ref{eq:Pkab}) we get

\begin{equation}
  \begin{aligned}
    \frac{(-1)^{d}}{16\pi} \delta \int_{\mathcal{BH}} \star (e^{a}\wedge
    e^{b}) P_{k\, ab}
    & = \frac{(-1)^{d}\kappa}{16\pi} \delta
    \int_{\mathcal{BH}} \star (e^{a}\wedge e^{b}) n_{ab}
    \\
    & \\
    & =
    -\frac{\kappa}{16\pi} \delta \int_{\mathcal{BH}} d^{d-2}S\, n_{ab}n^{ab}
    \\
    & \\
    & =
    T\delta A/4\, ,
    \\
  \end{aligned}
\end{equation}

\noindent
where we have used the normalization of the binormal $n_{ab}n^{ab}=-2$, $A$ is
the area of the horizon and $T=\kappa/2\pi$ is the Hawking temperature.

Thus, we recover the first law of black hole mechanics if we identify the
black hole entropy with one quarter of the area of the horizon.

%%%%%%%%%%%%%%%%%%%%%%%%%%%%%%%%%%%%%%%%%%%%%%%%%%%%%%%%%%%%%%%%%%%%%%
%%%%%%%%%%%%%%%%%%%%%%%%%%%%%%%%%%%%%%%%%%%%%%%%%%%%%%%%%%%%%%%%%%%%%%
%%%%%%%%%%%%%%%%%%%%%%%%%%%%%%%%%%%%%%%%%%%%%%%%%%%%%%%%%%%%%%%%%%%%%%
%%%%%%%%%%%%%%%%%%%%%%%%%%%%%%%%%%%%%%%%%%%%%%%%%%%%%%%%%%%%%%%%%%%%%%
\section{Discussion}
\label{sec-discussion}
%%%%%%%%%%%%%%%%%%%%%%%%%%%%%%%%%%%%%%%%%%%%%%%%%%%%%%%%%%%%%%%%%%%%%%
%%%%%%%%%%%%%%%%%%%%%%%%%%%%%%%%%%%%%%%%%%%%%%%%%%%%%%%%%%%%%%%%%%%%%%
%%%%%%%%%%%%%%%%%%%%%%%%%%%%%%%%%%%%%%%%%%%%%%%%%%%%%%%%%%%%%%%%%%%%%%
%%%%%%%%%%%%%%%%%%%%%%%%%%%%%%%%%%%%%%%%%%%%%%%%%%%%%%%%%%%%%%%%%%%%%%

In this paper we have showed how to define gauge-covariant Lie derivatives
with the momentum map and how to use these derivatives in the proof of the
first law of black-hole mechanics in the simple case of the Einstein-Maxwell
theory with the Vielbein as the gravitational field. We have also shown that
the momentum maps we have introduced in this case satisfy (well known) zeroth
laws. 

While the formulation of the first law of black-hole mechanics in the
Einstein-Maxwell theory is certainly not new, our proposal for dealing with
fields with gauge freedoms is a first step towards a generalization of the
first law to more complex cases involving $p$-form fields with Chern-Simons
terms such as those occurring in the Heterotic Superstring effective
action. Work in this direction is in progress \cite{kn:EMO}.

%%%%%%%%%%%%%%%%%%%%%%%%%%%%%%%%%%%%%%%%%%%%%%%%%%%%%%%%%%%%%%%%%%%%%%
%%%%%%%%%%%%%%%%%%%%%%%%%%%%%%%%%%%%%%%%%%%%%%%%%%%%%%%%%%%%%%%%%%%%%%
%%%%%%%%%%%%%%%%%%%%%%%%%%%%%%%%%%%%%%%%%%%%%%%%%%%%%%%%%%%%%%%%%%%%%%
%%%%%%%%%%%%%%%%%%%%%%%%%%%%%%%%%%%%%%%%%%%%%%%%%%%%%%%%%%%%%%%%%%%%%%
\section*{Acknowledgments}
%%%%%%%%%%%%%%%%%%%%%%%%%%%%%%%%%%%%%%%%%%%%%%b%%%%%%%%%%%%%%%%%%%%%%%%
%%%%%%%%%%%%%%%%%%%%%%%%%%%%%%%%%%%%%%%%%%%%%%%%%%%%%%%%%%%%%%%%%%%%%%
%%%%%%%%%%%%%%%%%%%%%%%%%%%%%%%%%%%%%%%%%%%%%%%%%%%%%%%%%%%%%%%%%%%%%%
%%%%%%%%%%%%%%%%%%%%%%%%%%%%%%%%%%%%%%%%%%%%%%%%%%%%%%%%%%%%%%%%%%%%%%

This work has been supported in part by the MCIU, AEI, FEDER (UE) grants
PGC2018-095205-B-I00 and PGC2018-096894-B-100, by the Spanish Research Agency
(Agencia Estatal de Investigaci\'on) through the grant IFT Centro de
Excelencia Severo Ochoa SEV-2016-0597 and by the Principau D'Asturies grant
FC-GRUPIN-IDI/2018/000174. The work of ZE has also received funding from ``la
Caixa'' Foundation (ID 100010434), under the agreement
LCF/BQ/DI18/11660042. TO wishes to thank M.M.~Fern\'andez for her permanent
support.

%%%%%%%%%%%%%%%%%%%%%%%%%%%%%%%%%%%%%%%%%%%%%%%%%%%%%%%%%%%%%%%%%%%%%%
%%%%%%%%%%%%%%%%%%%%%%%%%%%%%%%%%%%%%%%%%%%%%%%%%%%%%%%%%%%%%%%%%%%%%%
%%%%%%%%%%%%%%%%%%%%%%%%%%%%%%%%%%%%%%%%%%%%%%%%%%%%%%%%%%%%%%%%%%%%%%
%%%%%%%%%%%%%%%%%%%%%%%%%%%%%%%%%%%%%%%%%%%%%%%%%%%%%%%%%%%%%%%%%%%%%%
\appendix
%%%%%%%%%%%%%%%%%%%%%%%%%%%%%%%%%%%%%%%%%%%%%%%%%%%%%%%%%%%%%%%%%%%%%%
%%%%%%%%%%%%%%%%%%%%%%%%%%%%%%%%%%%%%%%%%%%%%%%%%%%%%%%%%%%%%%%%%%%%%%
%%%%%%%%%%%%%%%%%%%%%%%%%%%%%%%%%%%%%%%%%%%%%%%%%%%%%%%%%%%%%%%%%%%%%%
%%%%%%%%%%%%%%%%%%%%%%%%%%%%%%%%%%%%%%%%%%%%%%%%%%%%%%%%%%%%%%%%%%%%%%

%%%%%%%%%%%%%%%%%%%%%%%%%%%%%%%%%%%%%%%%%%%%%%%%%%%%%%%%%%%%%%%%%%%%%%
%%%%%%%%%%%%%%%%%%%%%%%%%%%%%%%%%%%%%%%%%%%%%%%%%%%%%%%%%%%%%%%%%%%%%%
%%%%%%%%%%%%%%%%%%%%%%%%%%%%%%%%%%%%%%%%%%%%%%%%%%%%%%%%%%%%%%%%%%%%%%
%%%%%%%%%%%%%%%%%%%%%%%%%%%%%%%%%%%%%%%%%%%%%%%%%%%%%%%%%%%%%%%%%%%%%%

\end{document}